**An Integrated Model for Financial Risk Assessment of Grid-ignited Wildfires**

**S. NEMATSHAHI[1*], A. KHODAEI[1], A. ARABNYA[1,2]**
[1]**University of Denver**
[2]**Quanta Technology**

**SUMMARY**

In recent years, the frequency and intensity of grid-ignited wildfires have increased significantly, leading to an elevated level of risk exposure to public safety and financial repercussions for electric utilities threatening their solvency. It is, therefore, imperative for electric utilities to accurately assess the financial impact of potential wildfires ignited by their power infrastructure. This is a critical step toward developing risk-informed strategies to mitigate grid-ignited wildfires from both operational and financial perspectives. This paper proposes and develops an integrated model to evaluate the damage costs associated with potential grid-ignited wildfires to allow assessing financial risk with greater precision than existing literature.

The proposed model is tailored to assess the financial risk associated with grid-ignited wildfires, including environmental damages, destroyed structures, and damage to the power grid assets. We quantify the risk associated with each power line, thereby identifying areas that require immediate preemptive actions. To visually represent the risk levels associated with the transmission grid topology, we implement a color-coded risk heatmap. The heatmap categorizes risk levels as follows: low-risk areas are denoted in white, moderate-low risk regions in green, medium-risk areas in yellow, and high-risk zones in red.

A reliable risk assessment provides several benefits for electric utilities. First, it enables them to identify High-Fire Risk Areas (HFRAs) that require immediate vegetation management and prioritize sections for undergrounding and conductor upgrades. By incorporating real-time weather data, electric utilities can make data-driven, risk-informed decisions to enhance their preemptive de-energization. Second, it can enhance situational awareness and early warning systems by identifying high-risk areas suitable for fire detectors and optimize the placement of Distributed Energy Resources (DERs) near these zones. Third, precise wildfire risk assessment enhances financial protection by promoting information exchange between electric utilities and insurance companies. This collaboration helps distribute the financial risk more effectively. By adopting the proposed approach, electric utilities can enhance operational efficiency, improve safety measures, and develop robust financial strategies to mitigate the impact of wildfires triggered by their power lines.

---

* Corresponding Author: S. Nematshahi (Member ID: 920240864)
  Ph.D. Candidate and Research Assistant (saeed.nematshahi@du.edu).    

## I. INTRODUCTION

Wildfires have the potential to cause extensive damage, including the burning of forests, the release of significant amounts of carbon dioxide, and the destruction of structures. At the same time, the impact of climate change combined with population growth in Wildland-Urban Interface (WUI) areas has led to increased frequency and intensity of wildfires [1]. For instance, rising temperatures, drastic changes in precipitation patterns, and extreme winds have intensified the wildfire drivers, making the situation more susceptible to fire ignition and spread. Fig. 1 illustrates the total acreage of burned area and the average burned area per wildfire in the U.S. between 1983 to 2023 [2]. As shown, the intensity of wildfires has increased over the last four decades [3].

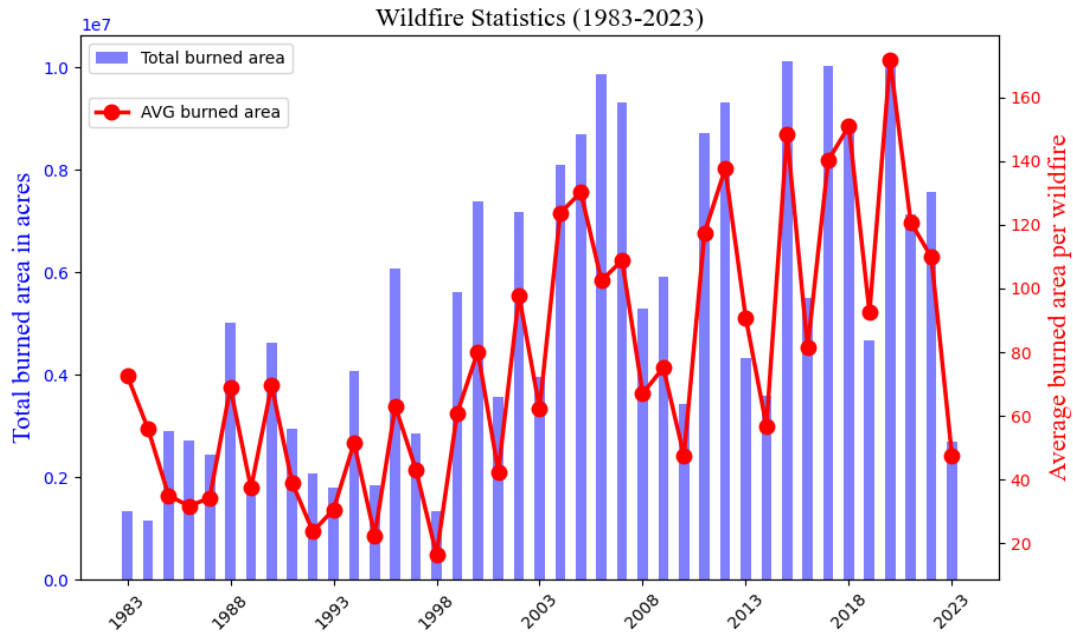

Fig. 1. Number of wildfires across the U.S. from 1983 to 2023.

Grid-ignited wildfires present a triple threat by causing extensive power outages, significant structural damages, and public safety risks. The 2024 Smokehouse Creek Fire in Texas and Oklahoma, for instance, became the largest wildfire in both states' histories, burning over one million acres and destroying hundreds of buildings. Similarly, California has experienced devastating wildfires. The 2021 Dixie Fire destroyed 1,329 structures, while the 2017 Tubbs Fire resulted in the loss of 5,636 structures [4]. The most tragic of them all was the 2018 Camp Fire, which claimed 85 lives and destroyed 18,804 structures. Authors in [5], stated that power lines were six times more prevalent in wildfires that destroyed houses than those where no houses were affected. Since the beginning of 2024, over 3,500 wildfires have occurred across California, resulting in the burning of approximately 207,000 acres of land by early July. This wildfire season in California has seen a dramatic increase, with about 20 times more acres burned compared to the same period last year [6]. Similarly, the financial losses from such large wildfires are considerably heightened and more challenging to recoup.

Multiple strategies exist to mitigate grid-ignited wildfires, such as undergrounding power cables in [7], probabilistic generation redispatch using MDP to minimize wildfire impacts on grid resilience in [8], and optimized power line de-energization to reduce wildfire risk and outage costs in [9]. However all require a reliable risk assessment, demanding further research. For electric utilities, conducting such a risk assessment is crucial to evaluate potential financial damages to third-party properties, the environment, and the power grid.

Among all causes of wildfire ignition, those ignited by power lines lead to the most financial loss, possibly even bankrupting electric utilities recognized as responsible for ignition. It is worth mentioning that the four costliest wildfires in U.S. history were all grid-ignited wildfires [10], and these four wildfires resulted in $35 billion in damages and about half a million people evacuated. This indicates a pressing concern regarding wildfires ignited by power lines. Fig. 2 depicts the frequency of grid-ignited wildfires in California from 2008 to 2022 [11]. The data indicates a substantial upward trend, needing further investigation into the underlying causes. When electric utilities are identified liable for the fire, they may be held liable for three principal categories of losses: property damage, suppression costs incurred by government agencies such as the United States Forest Service (USFS), and additional economic and natural resource damages [12]. The cost of suppressing large-scale wildfires is significantly lower compared to the extensive property damages incurred. For example, the Camp Fire, the most

expensive wildfire worldwide, had suppression costs of $150 million, while the total damages amounted to over $16.5 billion [13]. The substantial economic loss is primarily attributed to the expansion of residential development in the WUI, which has led to increased property destruction [14].

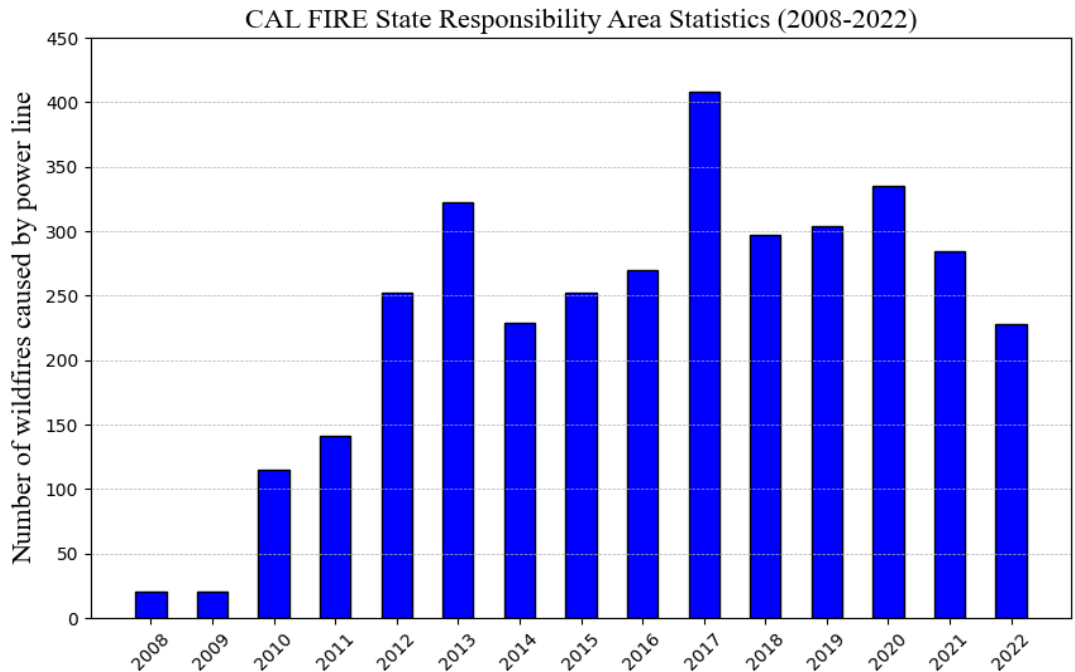

Fig. 2. Number of grid-ignited wildfires across California from 2008 to 2022.

Traditional wildfire risk assessment methods, such as the Fire Environment Mapping System (FEMS), are vital for long-term planning but often fail to account for specific ignition sources from power lines and don't consider the financial damage. Electric utilities need a practical wildfire risk assessment approach specifically tailored for fires ignited by power lines, incorporating an analysis of grid topology. Authors in [15] quantify the financial risk associated with each power line if it ignites a wildfire. They assumed a constant $20,000 damage cost per acre and stated that this amount highly depends on the number of structures, which is not considered in the existing literature. This omission highlights a research gap in understanding and assessing the risk of grid-ignited wildfires. Thus, we incorporated the actual data of third-party properties to appropriately estimate the total financial loss resulting from wildfires. In this paper, we propose an integrated model for financial risk assessment of grid-ignited wildfires. Our approach involves simulating wildfire scenarios ignited from power lines and calculating the total cost of damage to structures, the environment, and the power grid. This model uses actual data on landscape, weather, power grid infrastructure, and properties within the study area.

The subsequent sections of this paper are organized as follows: Section II delineates the methodology in detail. Section III presents the case study simulations and a discussion of the results. Finally, Section IV provides the conclusions and future work.

## II. METHODOLOGY

The proposed methodology for financial wildfire risk assessment involves five steps, including data integration, fire spread simulation, data wrangling, risk assessment, and risk mapping.

**Step 1: Data Integration**

The first step involves collecting necessary datasets for simulating grid-ignited wildfires and determining the corresponding cost of damages. This encompasses five sets of data as follows:

a) Landscape

The landscape file contains eight layers of information: elevation, slope, aspect, fuel model, canopy cover, stand height, canopy base height, and canopy bulk density. By determining the pixel size, these layers can be retrieved in a desired resolution. The landscape file is downloadable from LandFire [16], a program that provides data and tools for assessing and managing landscapes, particularly in the United States.

b) Weather data

The weather data is a record of the weather stream, including temperature, relative humidity, pressure, wind direction, and wind speed. The sampling interval is adjustable when downloading the weather data from the National Solar Radiation Database (NSRDB) [17].

c) Network topology

The network topology is represented by a GIS map that precisely delineates the locations of nodes and power lines. This grid map serves a dual purpose: to initiate the ignition points on power lines and to identify the affected lines as the wildfire spreads [18].

d) Ignition points

Before simulating a wildfire scenario, it is essential to establish ignition points. For grid-ignited wildfires, this can be carried out by evenly spaced points along each line of the transmission system. The precise coordinates of each ignition point are identified and recorded in an ignition file.

e) Structures map

A structure map delineates the precise locations of properties within the study area. The preliminary visualization is accessible via the OpenStreetMap (OSM) database [19], and the corresponding data is stored in an OSM file. This OSM data can be accessed and employed using a variety of tools and platforms, including QGIS (Quantum GIS), JOSM (Java OpenStreetMap Editor), and APIs such as the Overpass API, which has been utilized in this methodology to query specific data.

## Step 2: Wildfire Spread Simulation

The second step is simulating wildfire scenarios using FARSITE, a computer modeling system developed by the U.S. Forest Service for simulating the spread of wildfires across a landscape. It uses spatial information about topography, fuels, and weather to predict the behavior of a fire over time. The inputs and outputs for our wildfire simulation are explained as follows:

a) Inputs

It requires inputs including topography, fuel types, weather conditions, fuel moisture levels, fire behavior models, ignition points, and barriers. These inputs enable FARSITE to simulate and predict wildfire spread and behavior.

b) Outputs

FARSITE outputs include the fire perimeter, rate of spread, flame lengths, burn severity, fireline intensity, and heat per unit area. It also provides information on spot fire locations, helping to predict and understand wildfire behavior and its impacts.

Given the specified resolution, the study area encompasses a large number of pixels. The primary output of interest derived from the output file is the burned area, wherein each pixel is characterized by its unique row and column identifiers. The status of each pixel is denoted as 1 if it falls within the burned area, and as 0 if it does not.

## Step 3: Data Wrangling

The third step is identifying the affected communities. This methodology categorizes damages into three types: environmental damage, disruption to power lines requiring reconstruction, and damage to structures.

a) Environment

The burned environment is obtained by overlaying the fire perimeter onto the topographic map. The wildfire's extent in acres is then computed by multiplying the number of affected pixels by the pixel size.

b) Affected lines

To identify the lines impacted by the wildfire, the fire perimeter is overlayed onto the network topology map. A line segment is deemed affected by the wildfire if there is an intersection between the pixels representing the burned area and those associated with the line segment on the map.

b) Structures

To assess the number of structures destroyed due to wildfire spread, we first calculate the number of structures located within the burned area. This process entails creating a matrix that mirrors the dimensions of the FARSITE output, with each cell representing a count of structures. Subsequently, this matrix is overlaid onto the burned area matrix. The total number of destroyed structures is then determined by summing the values in the cells of the structures matrix that correspond to cells with a burned status of 1 in the burned area matrix.

At this stage, we have obtained key data points related to the wildfire's impact. Specifically, we have determined the total acres of the burned environment, identified a list of affected lines' segments within the network topology that intersect with the burned area, and calculated the total number of structures destroyed by the wildfire. This information collectively offers a pervasive assessment of the wildfire's effects on both the environment and infrastructure.

**Step 4: Risk Assessment**

The fourth step is calculating the cost of damage, encompassing several components: the financial loss to the environment, the expense of replacing the affected power lines, and the liability associated with damage to third-party structures. To facilitate this, an average cost is determined for each acre of burned environment, each mile of affected power lines, and each destroyed structure. This cost assessment is crucial for understanding the full economic impact of the wildfire. Finally, the total financial loss from the wildfire is obtained by aggregating these individual cost components.

**Step 5: Risk mapping**

The fifth step involves visualizing the network topology using a color-coded heatmap to represent the financial risk of potential wildfires on each power line. Power lines above the 90th percentile of the calculated risk will be highlighted in red, indicating a critical need for immediate mitigation measures. Lines in the 80th to 90th percentile range will be depicted in yellow, suggesting that these are priority areas for which risk mitigation is strongly recommended. Power lines falling between the 50th and 80th percentiles will be shown in green, identifying them as candidates for preemptive actions. Lines below the 50th percentile will be marked in white, signifying that they do not currently necessitate any immediate or planned interventions.

### III. NUMERICAL SIMULATIONS

#### A. Case study

The IEEE 30-bus system is selected as the representative transmission network for this study. This system includes six generators and 41 branches. The topology of the IEEE 30-bus system is mapped onto the geographical area of interest [20]. The study area is delineated by latitude bounds of 37.6° to 38.1° and longitude bounds of -120° to -120.7°. The actual weather data, starting from July 1st, 2022, is obtained to serve as a representative sample of summer conditions for the area under study. Ignition points are evenly distributed at 5-mile intervals throughout each power line. Fig. 3 illustrates this distribution, with red tags indicating ignition points and black lines representing the power lines.

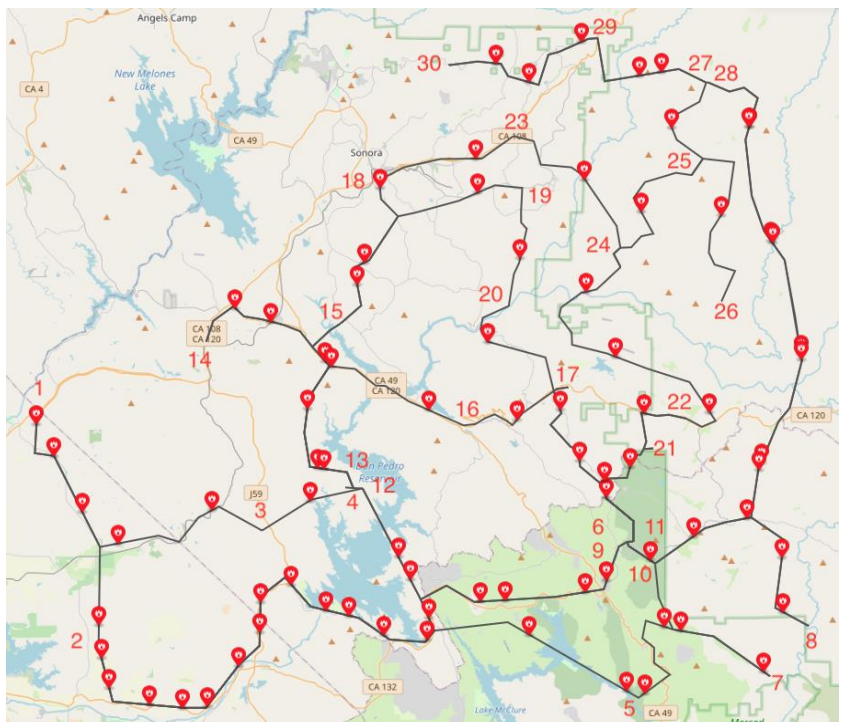

Fig. 3. Distribution of ignition points along power lines.

Using the available API, the number and location of structures within the area of study are extracted. This information is illustrated in a 2 by 2 cells figure to depict the density of properties. In Fig. 4, red dots indicate the locations of structures, while blue numbers inside the squares represent the number of structures within each grid cell. Note that the 2 by 2 grid depicted here is for illustrative purposes. In the simulation, a finer resolution is employed with a grid of 464 rows by 517 columns, resulting in 239,888 pixels, each representing an area of 120 meters by 120 meters.

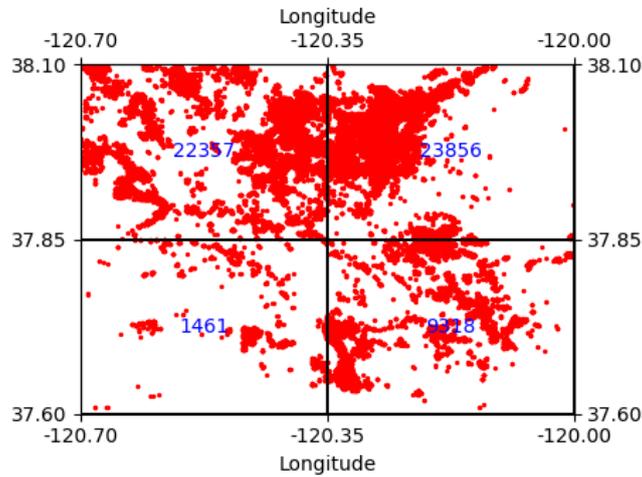

Fig. 4. Density of structures within the area of study.

At this stage, 66 scenarios, each with a single ignition point, are executed. The outputs from FARSITE are expressed in 464 by 517 grids. The burned area matrix is created by assigning a value of 1 to cells inside the fire zone, and a value of 0 to cells out of the fire zone. This burned area matrix determines the damages caused by the wildfire.

## B. Wildfire damages

To quantify the burned area in acres, we aggregate the number of cells in the burned area matrix that have a status of 1. This total is multiplied by a conversion factor of 3.559 to obtain the final measurement in acres. Fig. 5 presents a comparative analysis of the burned area, measured in acres, across 66 grid-ignited wildfire scenarios. It reveals that scenario 65, which ignites on line 41 (bus 6 to 28), and scenario 58 on line 40 (bus 8 to 28), exhibit the most extensive wildfire spread, whereas scenario 30 on line 10 (bus 6 to 8) and scenario 37 on line 22 (bus 27 to 29) are associated with the smallest burned areas.

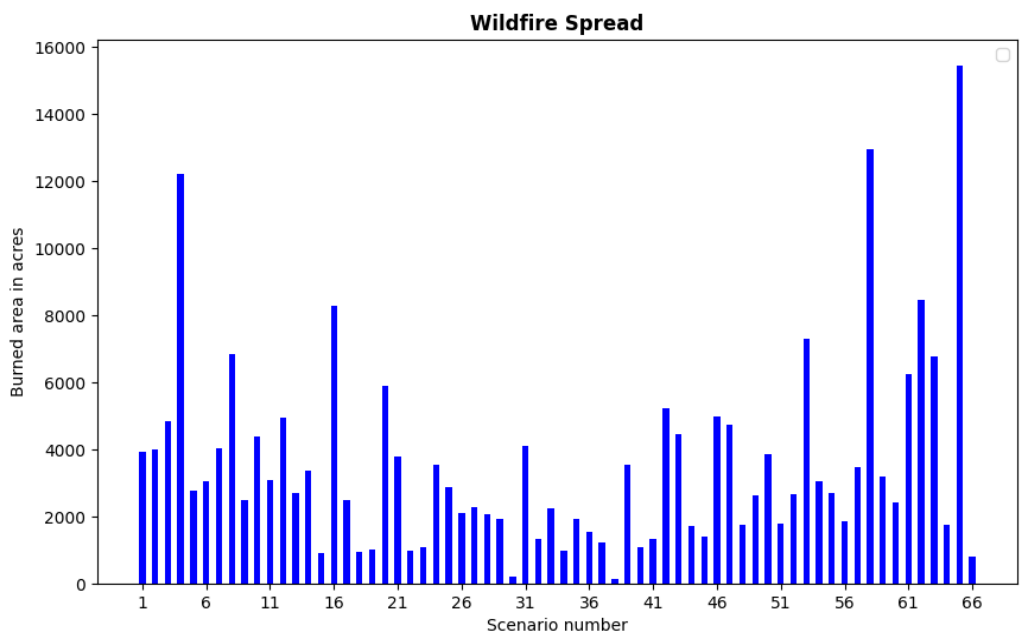

Fig. 5. Burned area as a result of grid-ignited wildfire scenarios.

To assess the extent of power lines requiring reconstruction, it is critical to integrate the network topology with the wildfire-affected area. In this study, power lines are subdivided into 1-mile segments. Any segment impacted by the wildfire mandates a complete 1-mile reconstruction. If all segments of a power line are affected, the entire length of the line is designated for replacement. Conversely, if only some segments are impacted, only the affected portions are considered for replacement. Fig. 6 depicts the length of power lines necessitating reconstruction under the grid-ignited wildfire scenarios. It indicates that scenario 3, on line 2 (bus 1 to 3), and scenario 58, on line 40 (bus 8 to 28), exhibit the greatest length of power lines requiring reconstruction, with each spanning 14 miles of transmission grid infrastructure.

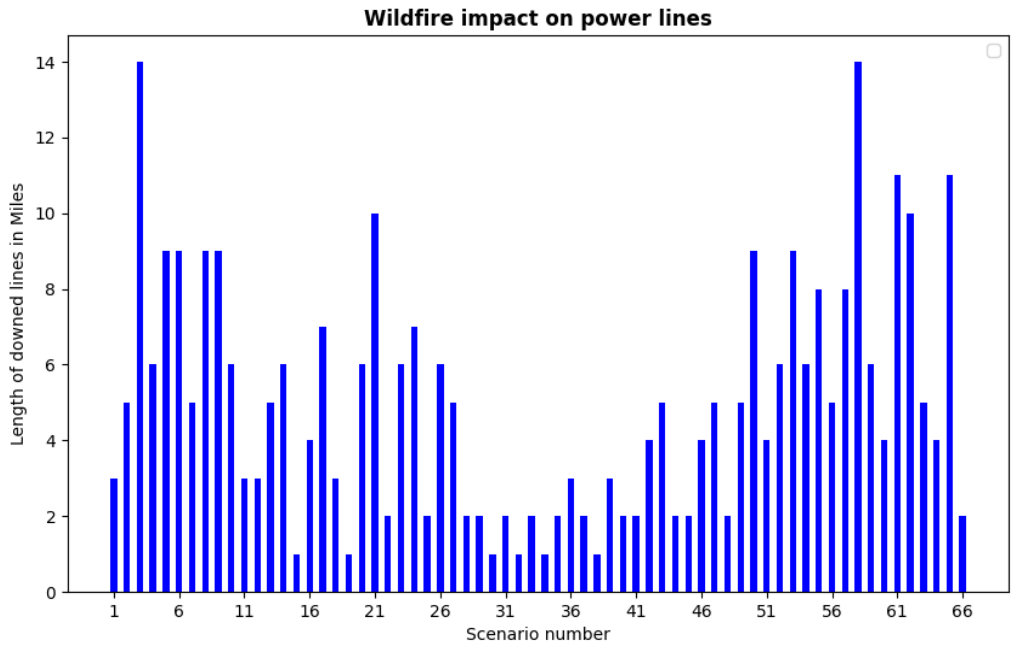

Fig. 6. Power lines affected by grid-ignited wildfire scenarios.

To evaluate the scope of damage to third-party structures in a wildfire scenario, we aggregate the number of structures located within pixels classified with a status of 1 in the burned area matrix. Adding up these counts, we obtain the total count of destroyed structures for each grid-ignited wildfire scenario. Fig. 7 illustrates the number of third-party structures destroyed due to the grid-ignited wildfire scenario. It demonstrates that, for certain scenarios, the number of destroyed structures is zero, suggesting that the burned area is situated in a wildland devoid of any buildings. Notably, the highest number of destroyed structures is observed in scenario 57, where the ignition point located on line 39, between bus 29 and bus 30, results in the destruction of 419 structures.

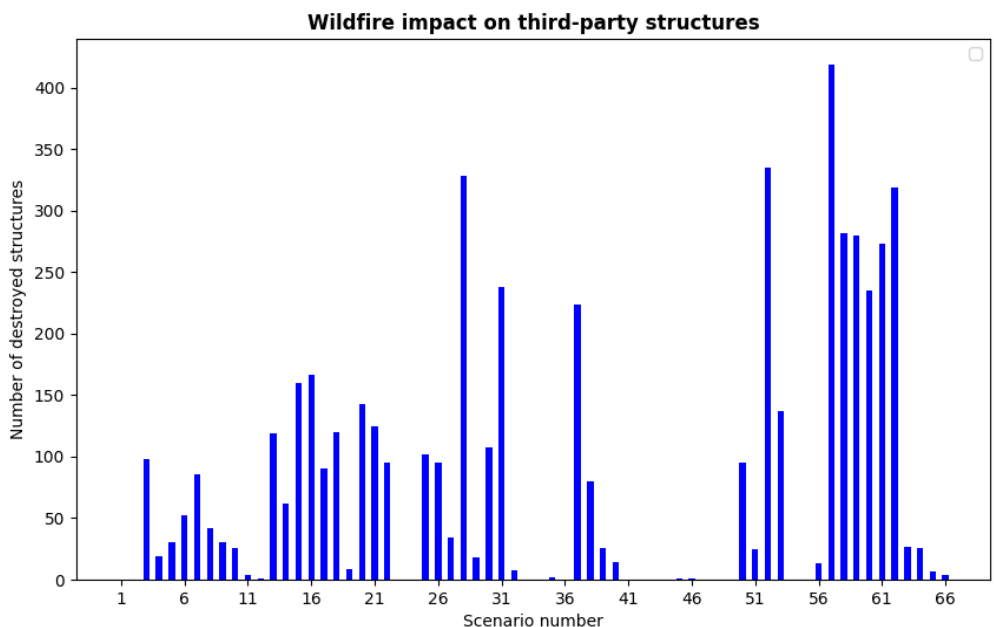

Fig. 7. Number of destroyed structures caused by grid-ignited wildfire scenarios.

## C. Cost of damages

The total financial loss encompasses three components: environmental damage, damage to power infrastructure, and damage to structures. For environmental damage, we estimate a cost of $500 per acre. The reconstruction cost for power lines is assumed to be $250,000 per mile. Additionally, we use the median home price in California as of April 2024, which is $904,210 [21], to represent the cost of damage to residential properties. The total financial loss from the grid-ignited wildfire is calculated through (1).

$$\begin{aligned}Total\ financial\ loss =\ & burned\ area\ (acres) * avg\ cost\ of\ environment\ damage\ (\$/acre) \\ & + length\ of\ downed\ power\ lines * avg\ cost\ of\ power\ line\ reconstruction\ (\$/mile) \\ & + number\ of\ destroyed\ structures * median\ home\ price\end{aligned} \quad (1)$$

Accordingly, Fig. 8 presents the total financial risk associated with the studied grid-ignited wildfire scenarios. The costliest wildfire is scenario 57, which is ignited from line 39 (connecting bus 29 to bus 30). While the burned area in this scenario is approximately average compared to other scenarios, the mid-high length of downed power lines and, more critically, the significant number of destroyed structures contribute to the highest total wildfire cost. Upon examining the details presented in Fig. 7 and Fig. 8, it is evident that the seven scenarios with the highest costs correspond to those with the greatest number of structures. This observation is consistent with historical data from real-world structures, corroborating the empirical evidence. Conversely, when the number of destroyed structures is relatively low, the primary considerations become the costs associated with the power grid infrastructure itself and the cost of damage to the environment.

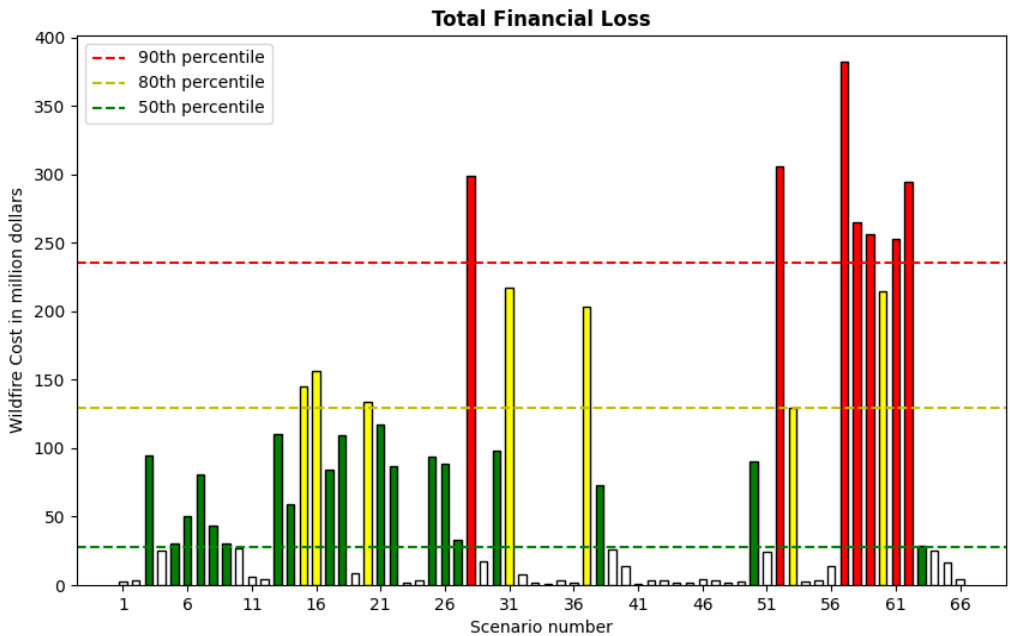

Fig. 8. Total cost resulted from grid-ignited wildfire scenarios.

## D. Risk heatmap

To make the results easy to follow, the final component involves visualizing areas of varying risk levels within the power grid. Fig. 9 illustrates the financial risk associated with power lines, with red lines indicating the highest risk, followed by yellow, green, and white lines, which represent progressively lower levels of risk. This visualization prioritizes areas for mitigation strategies by illustrating a color-coded network topology that enables electric utilities to plan for potential risks, with segments of power lines categorized by their wildfire ignition risk, ranging from highest to lowest.

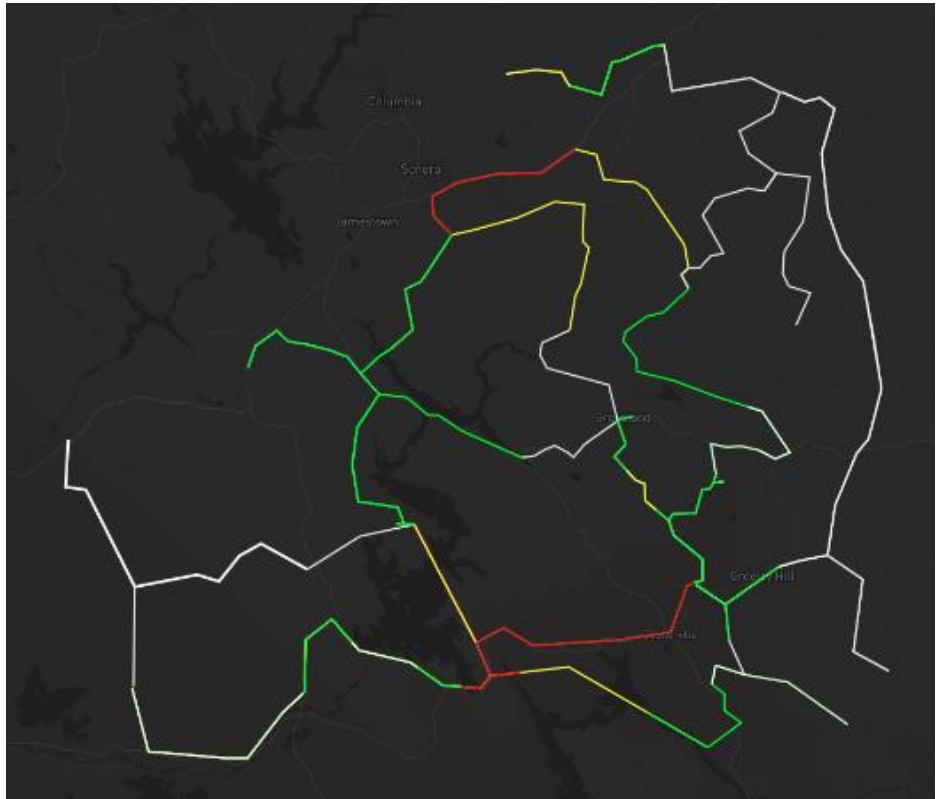
Fig. 9. Risk heatmap for the transmission network.

## IV. CONCLUSION

This study addresses the critical gap in financial risk assessments of grid-ignited wildfires by integrating data on the landscape, weather, power grid infrastructure, and structures within affected areas. The proposed integrated model for financial risk assessment of grid-ignited wildfires effectively simulates wildfire scenarios, calculates the resultant damage and corresponding financial losses, and visualizes the risk through a color-coded heatmap. The case study utilizing the IEEE 30-bus system demonstrates the model's ability to identify HFRA areas within the grid, providing a practical tool for electric utilities to prioritize wildfire mitigation efforts. By incorporating precise data and advanced simulation techniques, this model not only enhances the understanding of grid-ignited wildfire risks but also offers actionable insights for minimizing economic losses and improving wildfire management strategies.

The future research will focus on refining the model with more scenarios, including different seasons, wind speeds, and directions. Also, real-time data inputs could expand the model's application to the operational resilience of power systems against the escalating threat of wildfires.

## V. Disclaimer

# BIBLIOGRAPHY

[1] L. Johnston, R. Blanchi, M. Jappiot. (2019). Wildland-Urban Interface. In: Manzello, S. (eds) Encyclopedia of Wildfires and Wildland-Urban Interface (WUI) Fires. Springer, Cham. Available: https://doi.org/10.1007/978-3-319-51727-8_130-1.

[2] National Interagency Coordination Center, Source available: https://www.nifc.gov/fire-information/statistics/wildfires.

[3] Bayham, Jude, Jonathan K. Yoder, Patricia A. Champ, and David E. Calkin. "The economics of wildfire in the United States." *Annual Review of Resource Economics* 14, no. 1 (2022): 379-401.

[4] Lincoln Bramwell, "Understanding Wildfire in the Twenty-First Century: The Return of Disaster Fires," *Environmental History* 28, no. 3, 467-494, 2023.

[5] The National Fire Protection Association (NFPA). Wildland Fire Statistics. [Online]. Available: https://www.nfpa.org/Education-and-Research/Research/NFPA-Research/Fire-Statistical-reports/Wildland-Fire-Statistics [Accessed June 2024].

[6] Ayana Archie, "California's wildfires are burning far more land so far this year than in 2023," KCLU. [Online]. Available: https://www.kclu.org/environment/2024-07-16/californias-wildfires-are-burning-far-more-land-so-far-this-year-than-in-2023. [Accessed: July 20, 2024].

[7] S. Nematshahi, A. Khodaei and A. Arabnya, "An Investment Prioritization Model for Wildfire Risk Mitigation Through Power Line Undergrounding," 14th Mediterranean Conference on Power Generation, Transmission, Distribution and Energy Conversion (MEDPOWER 2024), Athens, Greece, 2024, pp. 1-6, in press.

[8] M. Abdelmalak and M. Benidris, "Enhancing Power System Operational Resilience Against Wildfires," in IEEE Transactions on Industry Applications, vol. 58, no. 2, pp. 1611-1621, March-April 2022, doi: 10.1109/TIA.2022.3145765.

[9] J. Su, S. Mehrani, P. Dehghanian and M. A. Lejeune, "Quasi Second-Order Stochastic Dominance Model for Balancing Wildfire Risks and Power Outages due to Proactive Public Safety De-Energizations," in IEEE Transactions on Power Systems, vol. 39, no. 2, pp. 2528-2542, March 2024, doi: 10.1109/TPWRS.2023.3289788.

[10] CalFire. https://www.fire.ca.gov/our-impact/statistics.

[11] Kousky, Carolyn, Katherine Greig, Brett Lingle, and K. Kunreuther. "Wildfire cost in California: The role of electric utilities." *changes* 114, no. 18 (2018): 4582-4590.

[12] K. M. Collins, T. D. Penman, and O. F. Price. "Some wildfire ignition causes pose more risk of destroying houses than others." *PLoS One* 11, no. 9 (2016): e0162083.

[13] A. Troy, J. Moghaddas, D. Schmidt, J. S. Romsos, D. B. Sapsis, W. Brewer and T. Moody, "An analysis of factors influencing structure loss resulting from the 2018 Camp Fire", *International Journal of Wildland Fire* 31, 586-598, 2022.

[14] A. Arab, A. Khodaei, R. Eskandarpour, M. P. Thompson and Y. Wei, "Three Lines of Defense for Wildfire Risk Management in Electric Power Grids: A Review," in IEEE Access, vol. 9, pp. 61577-61593, 2021, doi: 10.1109/ACCESS.2021.3074477.

[15] S. Nematshahi, A. Khodaei, A. Arabnya, "Risk assessment of transmission lines against grid-ignited wildfires," 2025 IEEE PES Grid Edge Technologies Conference & Exposition (Grid Edge), San Diego, CA, USA, 2025, pp. 1-5, in press.

[16] https://www.landfire.gov/

[17] National Solar Radiation Database. https://nsrdb.nrel.gov/

[18] S. Nematshahi, B. Sohrabi, A. Arabnya, A. Khodaei and E. Belval, "A Catastrophe Bond Design for the Financial Resilience of Electric Utilities Against Wildfires," in *IEEE Transactions on Energy Markets, Policy and Regulation*, doi: 10.1109/TEMPR.2024.3501012.

[19] https://www.openstreetmap.org/

[20] B. Sohrabi, A. Arabnya, M. P. Thompson and A. Khodaei, "A Wildfire Progression Simulation and Risk-Rating Methodology for Power Grid Infrastructure," in IEEE Access, doi: 10.1109/ACCESS.2024.3439724.

[21] T. Kurzweil, "California median home price passes $900K, sets all-time high," KTLA. [Online]. https://ktla.com/news/local-news/california-median-home-price-passes-900k-sets-all-time-high/. [Accessed: July 29, 2024].